\documentclass[aps,prb,reprint,amsmath,amssymb,superscriptaddress,showpacs,floatfix]{revtex4-2}

\usepackage{mathtools}
\usepackage[bb=dsserif]{mathalpha}
\usepackage{bm}
\usepackage{mleftright}
\mleftright

\usepackage[dvipsnames, table, rgb]{xcolor}
\definecolor{linkcolor}{rgb}{0.16, 0.32, 0.75}
\definecolor{citecolor}{rgb}{0.0, 0.5, 0.0}
\definecolor{urlcolor}{rgb}{0.06, 0.46, 1.0}
\usepackage[colorlinks, linkcolor={linkcolor}, citecolor={citecolor}, urlcolor={urlcolor}]{hyperref}

\usepackage{tikz}
\usetikzlibrary{3d,decorations.markings}

\begin{document}

\title{Effect of superconductivity on non-uniform magnetization in dirty SF junctions}

\author{A.\,V.~Levin}
\affiliation{Department of Physics, University of Ljubljana, Kongresni trg 12, 1000 Ljubljana, Slovenia}
\affiliation{Nanocenter CENN, Jamova 39, 1000 Ljubljana, Slovenia}
\author{P.\,M.~Ostrovsky}
\affiliation{Max Planck Institute for Solid State Research, 70569 Stuttgart, Germany}

\begin{abstract}
We study proximity effect in a tunnel junction between a bulk superconductor and a thin disordered ferromagnetic layer on its surface. Cooper pairs penetrating from the superconductor into the ferromagnet tend to destabilize its uniform magnetic order. The competition of this effect and the intrinsic magnetic stiffness of the ferromagnet leads to a second order phase transition between uniform and non-uniform magnetic states. Using the quasiclassical Usadel equation, we derive the Landau functional for this transition and construct the complete phase diagram of the effect. We identify a special point of ``resonance'' at which the characteristic energy scale of the proximity effect equals the exchange field of the ferromagnet. At this point, the uniform magnetic state is unstable even in the limit of large stiffness. We further explore the parameter regime far beyond the transition and determine the properties of the resulting strongly non-uniform magnetic state.
\end{abstract}

\maketitle

\section{Introduction}

Proximity effect arises in junctions between superconductors and normal metals and lays the foundation for numerous fascinating physical phenomena. The main microscopic mechanism behind proximity effect is the coherent penetration and propagation of Cooper pairs from the superconductor into the normal metal \cite{DeGennes1964}. One of the earliest and most studied manifestations of this mechanism is the Josephson effect \cite{Josephson1962} in the SNS junctions\cite{Golubov04, Likharev79}.

Mutual influence of the superconductors and normal metals becomes even more diverse in materials with magnetic order. A pristine ferromagnet is characterized by a uniform magnetization while Cooper pairs in a superconductor are in the singlet state of the electron spins. When the two materials are brought into contact (SF junction) Cooper pairs cannot easily penetrate into the ferromagnet due to the spin mismatch. Exchange field inside the ferromagnet converts a Cooper pair from the singlet state into a triplet with zero projection of the total spin on the direction of magnetization. However, Cooper pairs in both singlet and triplet spin states strongly decay in a uniform ferromagnet. The situation changes dramatically when the magnetic order in the ferromagnet is non-uniform, e.g.\ due to the presence of magnetic domains. Triplet Cooper pairs can propagate much easier in the magnetic material if the projection of their total spin on the magnetization direction is nonzero. This results in a long-range proximity effect in non-uniform (and noncollinear) ferromagnets---a topic of particular scientific interest in the last few decades \cite{BergeretEfetov05, Eschrig03, Klapwijk06, Halterman07, HouzetBuzdin07, Volkov08, Khaire10, Robinson10, Wang10, Sprungmann10, Anwar10, Kalenkov11, Leksin12, Banerjee14, Banerjee14-2, Mironov15, Martinez16, Glick18, Eskilt19, Vezin20, Stoddart22}.

In the present work, we take a different point of view on the same phenomenon. We will not specify any particular magnetic texture in the ferromagnetic part of the junction. Instead our goal is to find an optimal texture that will arise in a ferromagnet as a result of proximity to a superconductor. When singlet Cooper pairs enter the ferromagnet, electron spins participate in the exchange interaction with local magnetic moments. Since all such moments are naturally aligned in the same direction, exchange with Cooper pair electrons will inevitably lead to an attempt of spin flip. At the same time, the ferromagnet has its own magnetic stiffness that tends to align all localized moments in a single direction. Competition of these two effects may result in the phase transition between uniform and nonuniform magnetic states depending on the relative strength of proximity effect and ferromagnetic stiffness. To study the possibility of such a phase transition is the main objective of our work.

A similar problem about the phase transition into a nonuniform magnetic state in the SF junction was considered previously in Ref.\ \cite{BergeretEfetov00}. Main assumptions of that work are (i) temperature close to $T_c$ of the superconductor, (ii) transparent boundary conditions between S and F parts of the junction, and (iii) ballistic electron motion in the ferromagnet. These conditions result in a rather complicated theoretical model that includes both direct and inverse proximity effect and requires solving intricate equations for electron dynamics together with the self-consistency equation for the superconducting order parameter. The problem was only solved for a specific candidate nonuniform state---helical magnetic order---and the conclusion was that transition into such a state requires unrealistically strong proximity effect.

We will study a junction of the same geometry but under very different conditions. We assume a tunnel boundary between the superconductor and ferromagnet, diffusive electron motion, arbitrary temperature and allow for any nonuniform magnetic order. Taking advantage of the quasiclassical approximation ($E_F \tau \gg 1$, where $E_F$ is the Fermi energy and $\tau$ is the electron mean free time) \cite{Eilenberger68, Larkin69} and diffusive limit ($\Delta \tau \ll 1$, where $\Delta$ is the superconducting gap), we will describe the junction in terms of quasiclassical Green functions using the Usadel equation \cite{Usadel70, Altland98}. Tunnel junction between the superconductor and ferromagnet also allows us to neglect the inverse proximity effect on the superconducting side. We will demonstrate the possibility of the phase transition into a nonuniform magnetic state and establish a full phase diagram of the junction. We will also discuss the properties of the system well beyond the phase transition deep in the magnetically modulated phase.

The paper is organized as follows. In Sec.\ \ref{Sec:formalism}, we provide a detailed formulation of the problem and of the quasiclassical approximation. We also derive the Usadel equation suited for the hybrid structure with an arbitrary magnetic texture in the ferromagnetic part. In Sec. \ref{Sec:weak} we develop a perturbation theory for weakly nonuniform magnetic states in the Usadel equation. Within this perturbation theory, we derive the Landau functional describing the second order phase transition between uniform and nonuniform magnetization and develop a complete phase diagram of this effect. We also identify a specific point of ``resonance'' in parameter space where the proximity effect becomes very strong and perturbation theory fails. In Sec.\ \ref{Sec:helical}, we focus on a specific magnetic order---helical state---and consider the system for arbitrary parameters both near and far beyond the phase transition. We study magnetic modulation deep in the nonuniform state and also in the vicinity of the resonance. We summarize the results and discuss possible extensions of the model in Sec.\ \ref{Sec:summary}. Technically intricate calculations at the point of resonance are relegated to Appendix.

\section{Quasiclassical formalism}
\label{Sec:formalism}

\subsection{Statement of the problem}

\begin{figure}
    \includegraphics[width=0.7\columnwidth]{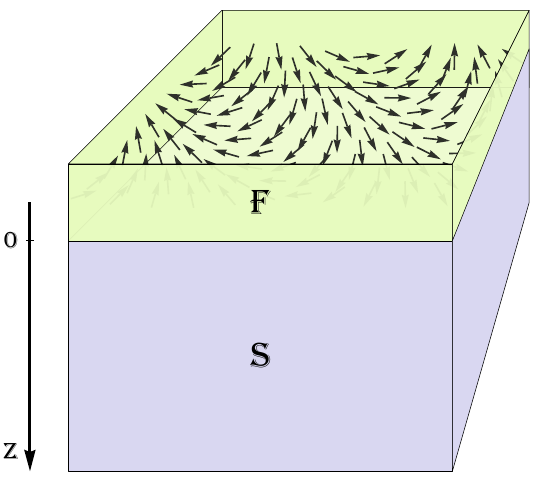}
    \caption{
    \label{fig:SF}
        Schematic depiction of the SF junction. A bulk superconductor (blue) is brought into contact with a thin ferromagnetic layer (green). Cooper pairs from the superconductor penetrate into the ferromagnet and can establish there a nonuniform magnetic order.}
\end{figure}

We consider a hybrid SF system consisting of a bulk superconductor and a ferromagnetic layer on its surface, see Fig.\ \ref{fig:SF}. The ferromagnetic layer will be treated as a two-dimensional metal with a local exchange field $\mathbf{h}$. This exchange field has a fixed absolute value but its direction can vary along the layer. Superconducting part of the junction is characterized by a constant value of the order parameter $\Delta$. Coupling between the superconductor and the ferromagnet is quantified by the interface conductance $G_T$ measured in units $e^2/\hbar$ per unit area.

The total free energy of the structure is a sum of magnetic and proximity contributions:
\begin{equation}
 F
  = F_\text{mag} + F_\text{prox}.
 \label{F}
\end{equation}
Magnetic part of the energy takes into account possible spatial variations of $\mathbf{h}$ and can be written as \footnote{
Here we use the notations $(\nabla \mathbf{n})^2 = \sum_{i,j} (\partial n_i/\partial x_j)^2$.} \cite{KuzminSkokov20}
\begin{equation}
 F_\text{mag}
  = \int d^2 r\; \frac{\zeta}{2} (\nabla \mathbf{n})^2.
 \label{Fmag}
\end{equation}
Here $\mathbf{n}$ is the unit vector in the direction of $\mathbf{h}$ and $\zeta$ is the 2D magnetic stiffness of the ferromagnet. Expression (\ref{Fmag}) is just the first term in the expansion of the free energy in the limit when the direction of magnetization slowly varies in space.

The second, proximity part of the free energy (\ref{F}) takes into account electron dynamics in the whole system. Electrons can penetrate the SF boundary and thus establish the proximity effect between superconductor and ferromagnet. We will treat this part of the free energy in the quasiclassical approximation. Quasiclassical Green function $g$ depends on a single point in real space, a single positive Matsubara energy $\epsilon$, and has the form of a $4 \times 4$ matrix in the Nambu-Gor'kov and spin spaces \cite{Altland98}. We will use notations $\tau_{x,y,z}$ and $\sigma_{x,y,z}$ for the Pauli matrices acting in these two spaces, respectively. In addition, quasiclassical Green function satisfies the constraint $g^2 = 1$ and hence can be represented as
\begin{equation}
 g
  = T^{-1} \tau_z T.
\end{equation}
Here $T \in \mathrm{U}(4)$ is any unitary matrix of size four. There is however a redundancy in associating $g$ and $T$. Two distinct matrices $T_{1,2}$ encode the same matrix $g$ if they differ by a left rotation $T_2 = K T_1$ with a unitary matrix $K$ that commutes with $\tau_z$. Such matrices $K$ form a subgroup $\mathrm{U}(2) \times \mathrm{U}(2)$ of the group $\mathrm{U}(4)$ of matrices $T$. Hence we can identify different matrices $g$ with right cosets of $K$ in $T$: $g \in \mathrm{U}(4) / \mathrm{U}(2) \times \mathrm{U}(2)$. This is a manifold of dimension eight hence the most general quasiclassical Green function can be parametrized by eight variables.

The quasiclassical Green function is determined by solving the Usadel equation \cite{Usadel70}. However, it is technically easier to work with the equivalent action
\begin{equation}
 S[g]
  = \pi\nu \int d^3r \operatorname{tr} \left[
      \frac{D}{4} (\nabla g)^2 - (\epsilon \tau_z + i \mathbf{h}\bm{\sigma} \tau_z + \Delta \tau_x) g
    \right].
 \label{Sg}
\end{equation}
Here $\nu$ is the density of states at the Fermi energy per one spin component and $D$ is the diffusion constant. In our setup, the action (\ref{Sg}) contains the term with $\Delta$ only inside the superconductor while the $\mathbf{h}$ term is present only inside the ferromagnetic layer. The values of $\nu$ and $D$ can be also different in the two materials.

The interface between superconducting and ferromagnetic parts of the junction is at $z = 0$, see Fig.\ \ref{fig:SF}. The Green function $g$ takes different values on the two sides of the interface, we will denote them $g(\pm 0)$. To take into account the coupling between two materials we add the following boundary term \cite{KuprianovLukichev88, Nazarov1994} to the action (\ref{Sg}):
\begin{equation}
 S_B
  = -\frac{\pi G_T}{4} \int d^2 r\,  \operatorname{tr} \Bigl[ g(+0) \: g(-0) \Bigr].
 \label{SB}
\end{equation}
Naturally, this part of the action involves only a 2D integral along the boundary, while Eq.\ (\ref{Sg}) contains a full volume integral.

The actual quasiclassical Green function minimizes the total action that is a sum of Eqs.\ (\ref{Sg}) and (\ref{SB}). Finding minimum of Eq.\ (\ref{Sg}) is equivalent to solving the Usadel equation while Eq.\ (\ref{SB}) provides proper boundary conditions. Once the Green function $g$ is found, it determines the minimized action $S_\text{min}(\epsilon)$ at a given Matsubara energy $\epsilon$. The free energy is then found as a sum over positive Matsubara energies
\begin{equation}
 F_\text{prox}
  = T \sum_{\epsilon > 0} S_\text{min}(\epsilon).
 \label{Fprox}
\end{equation}
As usual, fermionic Matsubara energies take odd discrete values $\epsilon = (2n + 1) \pi T$, where $T$ is the temperature. Our goal is to perform such a calculation for an arbitrary magnetic texture $\mathbf{h}$ and find the corresponding effective free energy (\ref{F}).

\subsection{Reduction of the Usadel equation}

We will now simplify quite general equations of the previous Section in several steps to take advantage of the geometry of our problem. Since we have a setup with only one bulk superconductor, we can fix the order parameter $\Delta$ to be real. As a consequence, the action (\ref{Sg}) contains only two Pauli matrices $\tau_x$ and $\tau_z$ in the Nambu-Gor'kov space. Matrix $g$, minimizing this action, will also contain only these two components. We can write it explicitly as
\begin{equation}
 g
  = \frac{1}{2} \Bigl[ (\tau_z - i \tau_x) Q + (\tau_z + i \tau_x) Q^{-1} \Bigr].
 \label{gQ}
\end{equation}
Here the matrix $Q$ operates only in the spin space and has the size $2 \times 2$. We have also chosen the form (\ref{gQ}) such that the condition $g^2 = \mathbb{1}$ is automatically satisfied.

The action (\ref{Sg}) takes the following form in terms of $Q$:
\begin{multline}
 S[Q]
  = \pi\nu \int d^3r \operatorname{tr} \biggl[
      \frac{D}{2} (\nabla Q) (\nabla Q^{-1})\\
      - (\epsilon + i \mathbf{h}\bm{\sigma} - i\Delta) Q - (\epsilon + i \mathbf{h}\bm{\sigma} + i\Delta) Q^{-1}
    \biggr].
 \label{SQ}
\end{multline}
The matrix $Q$ belongs to the unitary group $\mathrm{U}(2)$ with only four independent parameters in the spin space. These parameters are directly related to one singlet and three triplet components of the anomalous Green function \cite{IvanovFominov06}.

The boundary action (\ref{SB}) takes the following form with the ansatz (\ref{gQ}):
\begin{equation}
 S_B
  = -\frac{\pi G_T}{4} \int d^2 r\,  \operatorname{tr} \Bigl[ Q Q_S^{-1} + Q^{-1} Q_S \Bigr].
 \label{SBQ}
\end{equation}
Here $Q_S$ is the value of $Q$ in the superconductor right near the interface at $z = +0$ and $Q$ is the corresponding matrix on the ferromagnetic side $z = -0$. We will assume the coupling $G_T$ between superconductor and ferromagnet to be relatively weak such that the solution inside the superconductor is insensitive to the presence of the ferromagnet. In other words, the matrix $Q_S$ will be constant everywhere in the superconductor and will minimize the bulk action (\ref{SQ}) there disregarding the boundary term.

Inside the superconductor, exchange field $\mathbf{h}$ is absent and the action (\ref{SQ}) is fully rotationally symmetric. Its spatially uniform minimum $Q_S$ is a trivial matrix in the spin space and has the form
\begin{equation}
 Q_S
  = e^{i\theta_S},
 \qquad
 \tan\theta_S
  = \frac{\Delta}{\epsilon}.
 \label{thetaS}
\end{equation}
The angle $\theta_S$ is the standard Usadel angle that quantifies the conventional spin-singlet anomalous Green function. With this matrix $Q_S$, the boundary action (\ref{SBQ}) depends on the matrix $Q$ in the ferromagnetic layer only.

We further assume the ferromagnetic layer to be so thin, that the matrix $Q$ there, as well as the exchange field $\mathbf{h}$, does not change with $z$. Both quantities will be two-dimensional functions of lateral coordinates only. With these assumptions, the action (\ref{SQ}) in the ferromagnet together with the boundary contribution (\ref{SBQ}) becomes
\begin{multline}
 S[Q]
  = \pi\nu \int d^2r \operatorname{tr} \Bigl[
      \frac{D}{2} (\nabla Q) (\nabla Q^{-1})\\
      - (\epsilon + i \mathbf{h}\bm{\sigma} + \gamma e^{-i\theta_S}) Q - (\epsilon + i \mathbf{h}\bm{\sigma} + \gamma e^{i\theta_S}) Q^{-1}
    \Bigr].
 \label{SQh}
\end{multline}
Here we have the two-dimensional density of states $\nu$ that is proportional to the thickness of the ferromagnetic layer. We have also introduced an energy parameter
\begin{equation}
 \gamma
  = \frac{G_T}{4\nu}.
 \label{gamma}
\end{equation}
It appears due to the boundary term and provides the energy scale of the proximity effect. The most interesting interplay between ferromagnetism and proximity effect occurs when $\gamma$ is comparable to the exchange field in the ferromagnet. Let us note that in general the parameter $\gamma$ can be larger or smaller than the order parameter $\Delta$ in the superconductor. These  two cases have some qualitative differences that will be discussed later. For now we do not assume any relation between the energy scales $\gamma$ and $\Delta$.

We have thus established the effective two-dimensional action (\ref{SQh}) that is to be minimized with respect to $Q$ for a given configuration of the exchange field $\mathbf{h}$. We will further transform the action (\ref{SQh}) by applying a rotation in the spin space in order to align the field $\mathbf{h}$ along the $z$ axis. For any magnetic texture $\mathbf{h}$, the rotation is defined such that
\begin{equation}
 \mathbf{n} \bm{\sigma}
  = W^{-1} \sigma_z W,
 \qquad
 Q
  \mapsto W^{-1} Q W.
 \label{W}
\end{equation}
Here $\mathbf{n} = \mathbf{h}/h$ is the unit vector in the direction of $\mathbf{h}$ the same as in Eq.\ (\ref{Fmag}). This transformation replaces the term $\mathbf{h} \bm{\sigma}$ in the action (\ref{SQh}) with $h \sigma_z$. However, since the matrix $W$ varies in space, the gradients in the action will be replaced by long derivatives $\mathcal{D}$ defined as follows:
\begin{equation}
  \mathcal{D}
   = \nabla + i [\mathbf{A}, \cdot\;], \qquad
  \mathbf{A}
   = i (\nabla W) W^{-1}.
 \label{D}
\end{equation}
Physically, the vector $\mathbf{A}$ (whose components are spin matrices) plays the role of a nonabelian vector potential. It characterizes the rapidness of spatial rotation of the magnetization.

In the rotated frame, the action (\ref{SQh}) becomes
\begin{multline}
 S[Q]
  = \pi\nu \int d^2r \operatorname{tr} \biggl[
      \frac{D}{2} (\mathcal{D} Q) (\mathcal{D} Q^{-1}) \\
       -(\epsilon + i h \sigma_z + \gamma e^{-i\theta_S}) Q - (\epsilon + i h \sigma_z + \gamma e^{i\theta_S}) Q^{-1}
    \biggr].
 \label{SQA}
\end{multline}
All the information about nonuniform magnetization is now encoded inside the gradient term in the vector potential (\ref{D}). This action will be the main object of our further analysis.

\section{Perturbation theory}
\label{Sec:weak}

Throughout the paper we will assume that rotation of the magnetization in real space is relatively slow. This translates into relative smallness of the vector potential $\mathbf{A}$ defined in Eq.\ (\ref{D}). Our goal is to construct a series expansion $S_\text{min} = S_0 + S_1 + S_2 + \ldots$ of the minimized action (\ref{SQA}) in powers of $\mathbf{A}$. As will be shown below, this expansion goes in even powers, such that $S_1 \sim A^2$, $S_2 \sim A^4$ etc. The zero order term $S_0$ is independent of $A$ and hence will not be important for our analysis.

\subsection{Uniform state}

As the zero order approximation, we first neglect the vector potential altogether and consider the uniformly magnetized state. In this limit, the action (\ref{SQA}) has a complete translational invariance in the plane of the ferromagnetic layer and can be minimized by a constant matrix $Q_0$. The gradient term [first term in Eq.\ (\ref{SQA})] identically vanishes in this case. Since all the matrices explicitly entering the action (\ref{SQA}) are diagonal, we can also seek a solution in the diagonal form
\begin{equation}
 Q_0
  = \begin{pmatrix}
      e^{i\theta + m} & 0 \\
      0 & e^{i\theta - m}
    \end{pmatrix}.
 \label{Q0}
\end{equation}
Angles $\theta$ and $m$ can be readily found by minimizing Eq.\ (\ref{SQA}). Their values are such that
\begin{subequations}
\label{mtheta}
\begin{gather}
    \tanh(2m) = \frac{2h \gamma \sin \theta_{S}}{\epsilon^2 + h^2 + \gamma^2 + 2\epsilon \gamma \cos\theta_S}, \label{m} \\
    \tan(2\theta) = \frac{2\epsilon\gamma\sin\theta_S + \gamma^2 \sin(2\theta_S)}{\epsilon^2 + h^2 + 2\epsilon\gamma\cos\theta_S + \gamma^2 \cos(2\theta_S)}. \label{theta}
\end{gather}
\end{subequations}

Physical meaning of the angles $\theta$ and $m$ are quite simple. The angle $\theta$ is the standard Usadel angle appearing in the conventional Usadel equation. It quantifies the ratio of normal and anomalous parts of the local quasiclassical Green function and has the same meaning as the angle $\theta_S$ in Eq.\ (\ref{thetaS}) in the superconducting part of the junction. The parameter $m$ is analogous to $\theta$ but for the spin-triplet component of the quasiclassical Green function. In our case, magnetization is uniform and points in the $z$ direction hence the only triplet component is the one with zero projection of the total spin on the $z$ axis.

It should be noted that originally the matrix $Q$, as defined in Eq.\ (\ref{gQ}) belongs to the compact unitary group $\mathrm{U}(2)$. However, the matrix $Q_0$ from Eq.\ (\ref{Q0}) violates unitarity. This means that the uniform minimum of the action (\ref{SQA}) lies outside the original manifold where this action was derived. This is quite a standard situation for the saddle-point analysis: quasiclassical solution of a quantum mechanical problem involves trajectories that go into complex plane in terms of originally real coordinates. As a result, the matrix $Q_0$ provides a minimum of the action (\ref{SQA}) with respect to real deviations of $\theta$ and imaginary deviations of $m$ from the point (\ref{mtheta}).

\subsection{Weakly nonuniform state}

Once the magnetization $\mathbf{h}$ is nonuniform, the solution (\ref{Q0}) is no longer a true minimum of the action (\ref{SQA}). We assume that magnetization varies slowly in space and hence the vector potential $\mathbf{A}$ defined in Eq.\ (\ref{D}) is relatively small. This vector potential enters in the first, gradient term in Eq.\ (\ref{SQA}) inside the long derivatives. Hence we can treat the whole gradient term as a small perturbation. To the first order, the change of the action can be written as the value of this gradient term computed at the uniform solution $Q_0$:
\begin{multline}
 S_1
  = \frac{\pi\nu D}{2} \int d^2r \operatorname{tr} (\mathcal{D} Q_0) (\mathcal{D} Q_0^{-1}) \\
  = \frac{\pi\nu D}{2} \sinh^2 m \int d^2r \operatorname{tr} [\mathbf{A}, \sigma_z]^2.
 \label{SQ0}
\end{multline}
Here we have used the definition of the long derivative (\ref{D}).

The vector potential $\mathbf{A}$ is related to the rotation of the basis (\ref{W}) and hence to the direction of magnetization $\mathbf{n}$. These relations allow us to establish the identity
\begin{equation}
 \operatorname{tr} [\mathbf{A} ,\sigma_z]^2
  = -2 (\nabla\mathbf{n})^2.
\end{equation}
We see that the action (\ref{SQ0}) has exactly the same dependence on $\mathbf{n}$ as the proper magnetic free energy (\ref{Fmag}). Combining both contributions, we have the total free energy
\begin{equation}
 F
  = \int d^2r\, \frac{\zeta - \alpha}{2}\, (\nabla\mathbf{n})^2,
 \label{F2}
\end{equation}
with the parameter $\alpha$ given by the Matsubara sum
\begin{equation}
 \alpha
  = 2\pi\nu D T \sum_{\epsilon > 0} \sinh^2 m(\epsilon).
 \label{alpha}
\end{equation}
We observe that the proximity effect softens the ferromagnet by reducing its stiffness by $\alpha$. Once the difference $\zeta - \alpha$ is negative, the uniform state of the ferromagnet becomes unstable and the system undergoes a second order phase transition into a nonuniformly magnetized state.

In the remainder of this Section we will analyze the dependence of $\alpha$ on the three relevant energy scales of the system: magnetization magnitude $h$, superconducting order parameter $\Delta$, and coupling through the tunnel barrier $\gamma$. For simplicity, we will assume zero temperature limit and replace the Matsubara summation in Eq.\ (\ref{alpha}) by an equivalent integral. Analytic computation of this integral with $m$ from Eq.\ (\ref{m}) in its most general form still seems unfeasible. We will thus resort to two limiting cases of relatively weak ($\gamma \ll \Delta$) and strong ($\gamma \gg \Delta$) proximity coupling. In both limits, the parameter $\alpha$ is a function of the ratio $h/\gamma$.

\begin{figure}
    \centering
    \includegraphics[width=\columnwidth]{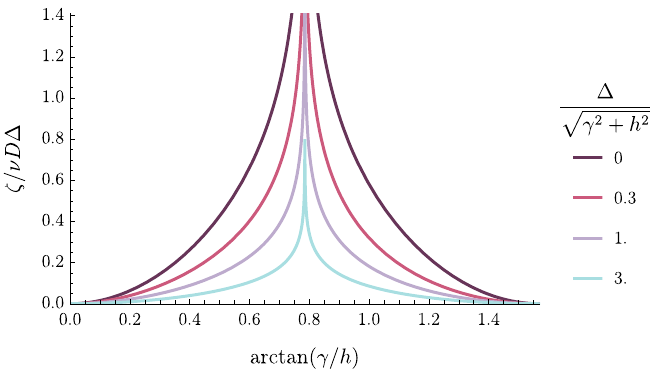}
    \caption{Phase diagram for the transition into a non-uniform magnetic state at zero temperature for different values of $\Delta$. Curves correspond to $\zeta = \alpha$ with $\alpha$ from Eq.\ (\ref{alpha}). The horizontal axis shows $\arctan(\gamma/h)$ in order to emphasize the symmetry $h \leftrightarrow \gamma$. On the vertical axis, the magnetic stiffness $\zeta$ is measured in units $\nu D \Delta$. The upper curve corresponds to Eq.\ (\ref{alphalargegamma}). The lowest curve approaches the limit of Eq.\ (\ref{alphasmallgamma}). Both limiting curves are symmetric under $\gamma \leftrightarrow h$ while the intermediate curves are very slightly asymmetric.
    }
    \label{fig::alpha delta}
\end{figure}

Consider first the case $\gamma \ll \Delta$. Relevant energies in Eq.\ (\ref{alpha}) are from the interval $0 < \epsilon \lesssim \gamma$. Hence, according to Eq.\ (\ref{thetaS}), we can replace $\theta_S = \pi/2$ in Eq.\ (\ref{m}). In the limit of zero temperature, Eq.\ (\ref{alpha}) reduces to a complete elliptic integral and yields
\begin{equation}
 \alpha
  = \nu D \begin{dcases}
      \gamma [K(h/\gamma) - E(h/\gamma)], & h < \gamma, \\
      h [K(\gamma/h) - E(\gamma/h)], & h > \gamma.
    \end{dcases}
 \label{alphasmallgamma}
\end{equation}
This function $\alpha$ is shown in Fig.\ \ref{fig::alpha delta}. It is completely symmetric with respect to $h \leftrightarrow \gamma$ and diverges at the symmetry point $h = \gamma$. This is physically the most interesting point. Diverging $\alpha$ means that no matter how stiff the ferromagnet is, the difference $\zeta - \alpha$ will inevitably become negative close to the point $h = \gamma$. We thus conclude that in the vicinity of this point the system should exhibit a transition to a nonuniform magnetic state irrespective of the value of $\zeta$. We will refer to this point in the parameter space as the resonance.

\begin{figure}
    \centering
    \includegraphics[width=\columnwidth]{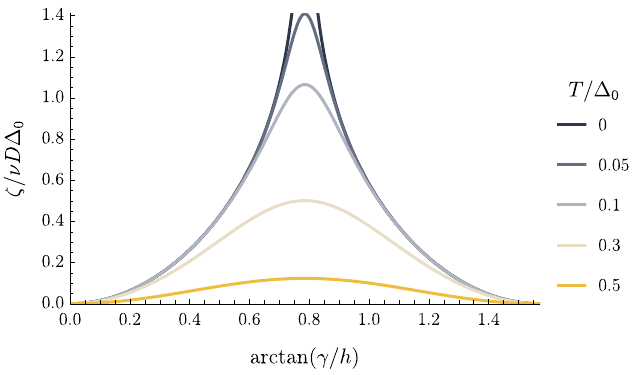}
    \caption{Phase diagram for the transition into a non-uniform magnetic state in the limit $\gamma \gg \Delta$ at different temperatures. Transition occurs at $\zeta = \alpha$ with the value of $\alpha$ according to Eq.\ (\ref{alpha}). At zero temperature, the curve corresponds to Eq.\ (\ref{alphalargegamma}) and diverges at the point of resonance $h = \gamma$. For non-zero temperatures we also take into account suppression of the order parameter $\Delta(T)$ according to the standard BCS theory \cite{deGennesBook99}. The horizontal axis is the same as in Fig.\ \ref{fig::alpha delta}. On the vertical axis, the magnetic stiffness $\zeta$ is measured in units $\nu D \Delta_0$ where $\Delta_0$ is the value of $\Delta$ at $T = 0$. Maximal possible value of $T$ is the critical temperature $T_C \approx 0.57 \Delta_0$.}
    \label{fig::alpha temperature}
\end{figure}

In the opposite limit of relatively strong proximity effect $\gamma \gg \Delta$, relevant energies in Eq.\ (\ref{alpha}) are from the interval $0 < \epsilon \lesssim \Delta$. We can neglect explicit appearance of $\epsilon$ in Eq.\ (\ref{m}) and retain the energy dependence only via the parameter $\theta_S$. We take the zero temperature limit and change integration variable from $\epsilon$ to $\theta_S$ according to Eq.\ (\ref{thetaS}). This will again bring Eq.\ (\ref{alpha}) to the form of a complete elliptic integral
\begin{equation}
 \alpha
  = \frac{\nu D \Delta}{2} \left[
      K\left( \frac{2 h \gamma}{\gamma^2 + h^2} \right) - E\left( \frac{2 h \gamma}{\gamma^2 + h^2} \right)
    \right].
 \label{alphalargegamma}
\end{equation}
This function is also shown in Fig.\ \ref{fig::alpha delta}. It is again symmetric under the interchange $h \leftrightarrow \gamma$ and diverges at the resonance $h = \gamma$. We also show the value of $\alpha$ at different nonzero temperatures in Fig.\ \ref{fig::alpha temperature}.

Vicinity of the resonance $h = \gamma$ is the most interesting part of our solution since the proximity effect is strongest there. We can analyze this parameter range for an arbitrary ratio $\gamma/\Delta$. Divergence happens due to contribution of small energies in Eq.\ (\ref{alpha}). Making an expansion of Eq.\ (\ref{m}) in small $\epsilon$ and in small difference $h - \gamma$, we can estimate
\begin{equation}
 \alpha
  \approx \frac{\nu D}{2} \frac{\gamma \Delta}{\gamma + \Delta} \ln \min\left\{
      \frac{\gamma}{|h - \gamma|},\, \frac{\Delta}{T},\, \frac{\gamma}{T}
    \right\}.
 \label{alphalog}
\end{equation}

We have thus established conditions for the second order phase transition between states with homogeneous and inhomogeneous magnetization. It happens when the parameter $\alpha$ exceeds the magnetic stiffness $\zeta$. In order to analyze the inhomogeneous state that emerges at the transition, we need to expand the free energy to a higher order in gradients of $\mathbf{n}$. This will be done in the next Section.

\subsection{Landau functional}

Once the parameter $\alpha$ exceeds $\zeta$ the ferromagnetic order becomes nonuniform. Close to the transition, the leading term of the free energy expansion (\ref{F2}) almost vanishes and the next term in gradients of $\mathbf{n}$ gets important. To derive this next term, we should also take into account deviations of $Q$ from the uniform solution (\ref{Q0}). We will quantify this deviation by a matrix $X$ and write an expansion near the uniform solution as
\begin{equation}
 Q
  = Q_0(1 + X + X^2/2 + \ldots).
 \label{QX}
\end{equation}

We treat the complete first term of the action (\ref{SQA}) as a small perturbation. This term is of the order $A^2$. Hence $X \sim A^2$ and we will expand the action to the second order in $X$ and to the fourth order in $A$. Performing this expansion and using the value of $Q_0$ from Eq.\ (\ref{Q0}), we obtain the second order term in the expansion of the action
\begin{gather}
 S_2
  = \pi\nu \int d^2r \operatorname{tr} \left[
      D X \mathcal{D} (Q_0^{-1} \mathcal{D} Q_0)
      -\begin{pmatrix} \lambda_+ & 0 \\ 0 & \lambda_- \end{pmatrix} X^2
    \right], \label{S2X} \\
 \lambda_\pm
  = \sqrt{\bigl( \epsilon \pm ih + \gamma e^{-i\theta_S} \bigr) \bigl( \epsilon \pm ih + \gamma e^{i\theta_S} \bigr)}.
\end{gather}
It remains to minimize this quadratic action with respect to $X$.

The first, linear term of Eq.\ (\ref{S2X}) contains the matrix $\mathcal{D} (Q_0^{-1} \mathcal{D} Q_0)$. This matrix has zero trace and hence can be expanded in three Pauli matrices in the spin space:
\begin{equation}
 \mathcal{D} (Q_0^{-1} \mathcal{D} Q_0)
  = \mathbf{c} \bm{\sigma}.
 \label{csigma}
\end{equation}
Using this expansion we can minimize the action (\ref{S2X}) explicitly in terms of components of $\mathbf{c}$ and obtain
\begin{equation}
 S_2
  = \pi \nu D^2 \int d^2r \left[
      \frac{c_x^2 + c_y^2}{\lambda_+ + \lambda_-} + \frac{c_z^2}{4} \left( \frac{1}{\lambda_+} + \frac{1}{\lambda_-} \right)
    \right].
 \label{S2}
\end{equation}
Elements of the vector $\mathbf{c}$ can be found directly by substituting $Q_0$ from Eq.\ (\ref{Q0}) into Eq.\ (\ref{csigma}). Combinations of vector components that enter Eq.\ (\ref{S2}) can be then related to the gradients of magnetization by using the definitions (\ref{W}) and (\ref{D}). The resulting expressions are
\begin{align}
 &c_z^2
  = \frac{\sinh^2(2m)}{8} \operatorname{tr} (\mathcal{D} \sigma_z)^4
  = \frac{\sinh^2(2m)}{4}\, (\nabla \mathbf{n})^4, \\
 &c_x^2 + c_y^2
  = \frac{\sinh^2 m}{2} \operatorname{tr} \left[ (\mathcal{D}^2 \sigma_z)^2 - (\mathcal{D} \sigma_z)^4 \right] \notag \\
  \MoveEqLeft[-11] = \sinh^2 m \left[ \mathbf{n} \times \nabla^2\mathbf{n} \right]^2.
\end{align}

With these values we can rewrite the action (\ref{S2}) directly in terms of the gradients of $\mathbf{n}$. Remaining summation over Matsubara energies in Eq.\ (\ref{Fprox}) provides the next term in the expansion of the free energy. Together with Eq.\ (\ref{F2}), this yields the following Landau functional for the phase transition between uniform and nonuniform states:
\begin{equation}
 F
  = \int d^2r \left( \frac{\zeta - \alpha}{2}\, (\nabla\mathbf{n})^2 + \frac{\beta}{4} (\nabla\mathbf{n})^4 + \frac{\beta'}{4} \left[ \mathbf{n} \times \nabla^2\mathbf{n} \right]^2 \right).
 \label{Landau}
\end{equation}
Two new coefficients in this expansion are defined as
\begin{align}
 \beta
  &= \pi \nu D^2 T \sum_{\epsilon > 0} \frac{\sinh^2(2m)}{4} \left( \frac{1}{\lambda_+} + \frac{1}{\lambda_-} \right), \label{beta} \\
 \beta'
  &= 4\pi \nu D^2 T \sum_{\epsilon > 0} \frac{\sinh^2 m}{\lambda_+ + \lambda_-}. \label{betaprime}
\end{align}

\begin{figure}
  \centering
    \includegraphics[width=\columnwidth]{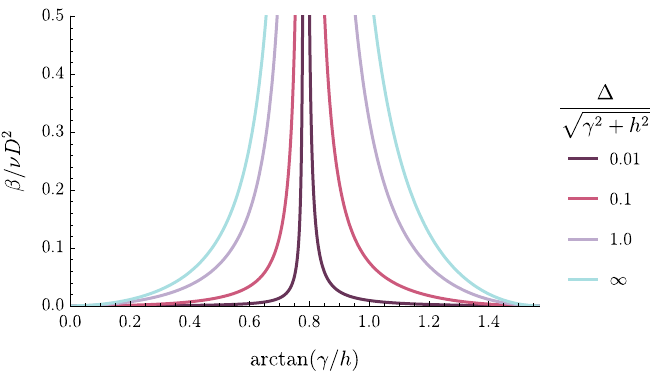}\\
    \includegraphics[width=\columnwidth]{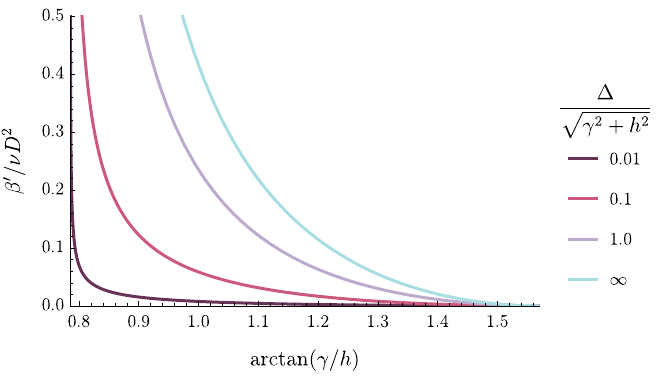}
    \caption{Dependence of $\beta$ [Eq.\ (\ref{beta}), upper panel] and $\beta'$ [Eq.\ (\ref{betaprime}), lower panel] on $\gamma/h$ for different values of $\Delta$ at zero temperature. Horizontal axis is the same as in Figs.\ \ref{fig::alpha delta} and \ref{fig::alpha temperature}. Parameter $\beta'$ diverges at $h > \gamma$ according to Eq.\ (\ref{beta' zero temperature divergences}) hence only half of the horizontal axis in the lower panel is shown.}
  \label{fig::beta delta}
\end{figure}

Similar to the parameter $\alpha$ discussed in the previous Section, the two extra parameters $\beta$ and $\beta'$ are functions of three energy scales: magnetization $h$, superconducting gap $\Delta$, and temperature $T$. Dependence of $\beta$ and $\beta'$ on the ratio $h/\gamma$ is shown in Fig.\ref{fig::beta delta} for several values of $\Delta/\gamma$. Unlike $\alpha$, there is no symmetry under the interchange $h \leftrightarrow \gamma$. Both parameters diverge at zero temperature at the resonance point $h = \gamma$ even stronger than $\alpha$. The parameter $\beta$ diverges as $|h - \gamma|^{-3/2}$ while $\beta' \sim |h - \gamma|^{-1/2}$. Moreover, $\beta'$ diverges logarithmically at zero temperatures everywhere in the region $h > \gamma$,
\begin{equation}
 \beta'
  \approx \frac{\nu D^2 \Delta \gamma^2}{(\Delta + \gamma) h \sqrt{h^2 - \gamma^2}} \ln \frac{\min\{\Delta, h\}}{T}.
  \label{beta' zero temperature divergences}
\end{equation}

We have thus established the full Landau functional (\ref{Landau}) for the second order phase transition into a nonuniform magnetic state. This functional takes into account proximity effects due to coupling to the superconductor. Any additional intrinsic or geometric effects in the ferromagnetic layer, like e.g.\ magnetic anisotropy or demagnetizing factor, can be also added to this functional directly. Particular magnetic texture arising at the transition can be then derived from the full functional. In the next Section, we will consider one possible solution to this problem.

\section{Helical state}
\label{Sec:helical}

\begin{figure}
    \centering
    \includegraphics[width=\columnwidth]{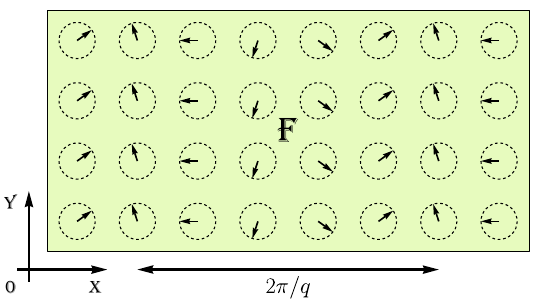}
    \caption{Illustration of the helical state in the ferromagnetic layer, view from above. Wave vector $q$ points to the right and the direction of magnetization rotates in the horizontal plane with the aperture angle $\psi = \pi/2$, cf.\ Eq.\ (\ref{n}). }
    \label{fig:Helical state demonstration}
\end{figure}

Minimizing Landau functional (\ref{Landau}) in its full form means solving a corresponding nonlinear differential equation. Among many possible nontrivial solutions there is always a particularly simple one---the helical state, in which magnetization rotates in real space with a constant rate. We will consider the helical state of the following form:
\begin{equation}
 \mathbf{n} 
 = \begin{pmatrix}
     \sin\psi \cos(qx)\\ \sin\psi \sin(qx) \\ \cos\psi
    \end{pmatrix}.
 \label{n}
\end{equation}
Magnetization rotates with the wave vector $q$ along a cone with the aperture $\psi$. We choose the $x$ axis in real space in the direction of $q$ and the $z$ axis in spin space along the axis of the cone, Fig.\ \ref{fig:Helical state demonstration}.

We can determine the values of $\psi$ and $q$ by minimizing the free energy. For the helical state (\ref{n}), spatial density of the Landau functional (\ref{Landau}) is constant and equals
\begin{equation}
 F
  = \frac{q^2}{2} \sin^2\psi \left[\zeta - \alpha + \frac{q^2}{2} \Bigl( \beta \sin^2\psi + \beta' \cos^2\psi \Bigr) \right].
 \label{Landaupsiq}
\end{equation}
This expression always has a trivial stationary point with $\psi = 0$ corresponding to the uniform state. However, when $\alpha$ exceeds $\zeta$ this stationary point becomes unstable and another solution is realized. Minimizing the free energy with respect to $\psi$ and $q$, we find the aperture $\psi = \pi/2$ and
\begin{equation}
   q =  \begin{dcases}
        0, & \alpha < \zeta,  \\
        \sqrt{\frac{\alpha - \zeta}{\beta}}, &\alpha > \zeta.
    \end{dcases}
    \label{FinalAnswerForQNextPhaseTransition} 
\end{equation}
We have thus identified the parameters of the helical magnetic state near the phase transition from the uniform state. Here $\alpha$ and $\beta$ are functions of $h$ and $\gamma$ as was discussed in the previous Section. Remarkably, the value of $\beta'$ turns out to be unimportant since the optimal helical state corresponds to the rotation of $\mathbf{n}$ along a great circle in the spin space with $\psi = \pi/2$.

Our model does not include any spin-orbit effects hence the orientation of $\mathbf{n}$ in spin space is unrelated to the orientation of the sample in real space. However, if magnetization rotates parallel to the plane of the sample, as shown in Fig.\ \ref{fig:Helical state demonstration}, it will further minimize the energy of the external magnetic field created outside the sample.

\subsection{Developed helical state}

Helical state provides a valid minimum of the action not only close to the phase transition but actually for any values of the parameters also deep inside the region of nonuniform magnetization on the phase diagram Fig.\ \ref{fig::alpha delta}. Far from the phase transition, the value of the wave vector $q$ is no longer small and we will refer to this solution as the developed helical state.

We can directly compute the action (\ref{SQA}) for the configuration (\ref{n}) without resorting to the perturbative expansion of Section \ref{Sec:weak}. An important advantage of the solution (\ref{n}) is that the action density is constant in space and equals
\begin{multline}
 S
  = -\pi\nu \biggl[
      D q^2 \sinh^2 m - 4 h \sin\theta \sinh m \\
      +4 \Bigl( \epsilon \cos\theta + \gamma \cos(\theta - \theta_S) \Bigr) \cosh m 
    \biggr].
 \label{Sdeveloped}
\end{multline}
In the uniform case $q = 0$, minimization of this action yields Eqs.\ (\ref{mtheta}) for $\theta$ and $m$. For a nonzero $q$, this minimization can be performed numerically. Subsequent summation over the Matsubara energies yields the proximity part of the free energy (\ref{Fprox}). Together with the magnetic contribution (\ref{Fmag}), we have the following free energy density for the helical state:
\begin{equation}
 F
  = \frac{\zeta q^2}{2} + T \sum_\epsilon S_\text{min}(\epsilon, q).
 \label{FV}
\end{equation}
This expression is to be minimized with respect to $q$ in order to find the optimal value of the wave vector in the helical state.

At the point of its minimum, the action (\ref{Sdeveloped}) is stationary with respect to variations in $m$ or $\theta$. Its only dependence on $q$ comes from the first term, where $q^2$ enters explicitly. Using this fact, we can take a derivative in $q$ of the total free energy density (\ref{FV}) and find the following general equation for the optimal value of $q$:
\begin{equation}
 \zeta
  = 2\pi\nu D T \sum_{\epsilon > 0} \sinh^2 m(\epsilon, q).
 \label{zetaq}
\end{equation}
Here $m(\epsilon,q)$ is the value of $m$ that minimizes the action (\ref{Sdeveloped}) at a given energy. Let us stress that Eq.\ (\ref{zetaq}) works equally well anywhere in the parameter space. Both near and far from the phase transition into the helical state. 

\begin{figure}
    \centering
    \includegraphics[width=\columnwidth]{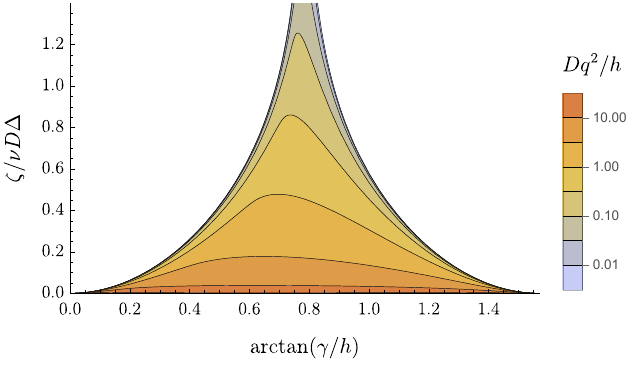}
    \caption{Dependence of $Dq^2/h$ (in color, logarithmic scale) for the helical state on $\arctan(\gamma/h)$ (horizontal axis) and $\zeta/\nu D \Delta$ (vertical axis). Near the phase transition (close to the border of the colored region), the amplitude of the wave vector $q$ obeys Eq.\ (\ref{FinalAnswerForQNextPhaseTransition}). Upper part of the plot near $h = \gamma$ corresponds to the resonance. Wave vector $q$ is given by Eq.\ (\ref{qresonance}) there. Developed helical state is at the bottom of the plot with small $\zeta$ and relatively large $q$, cf.\ Eq.\ (\ref{qdeveloped}).}
    \label{fig:helical state}
\end{figure}

We show the result of numerical solution of Eq.\ (\ref{zetaq}) in Fig.\ \ref{fig:helical state}. Boundaries of the nonuniform phase in this plot are the same as in Fig.\ \ref{fig::alpha delta}. The color inside the nonuniform region shows the optimal value of $q$. We observe that $q$ is indeed small close to the phase transition and then grows to relatively large values far from the transition.

In the limit of relatively small $\zeta$, lower part of the plot in Fig.\ \ref{fig:helical state}, the value of $q$ gets large. We can study this asymptotic regime analytically by assuming $m$ in Eq.\ (\ref{Sdeveloped}) to be small. In this limit, we can replace $\cosh m \approx 1$ in the action (\ref{Sdeveloped}) and find $\sinh m$ from the first two terms of the action:
\begin{equation}
 \sinh m
  \approx \frac{2h}{Dq^2} \sin\theta
  \approx \frac{2h}{Dq^2} \frac{\gamma \sin\theta_S}{\sqrt{\epsilon^2 + \gamma^2 + 2\epsilon\gamma\cos\theta_S}}.
\end{equation}
Optimal value of $\theta$ here is determined from the minimization of the last two terms of the action.

With this value of $m$, we can find the optimal $q$ from Eq.\ (\ref{zetaq}). In the two limiting cases of either $\gamma \gg \Delta$ or $\gamma \ll \Delta$ this value is
\begin{equation}
 q
  = \sqrt{\frac{h}{D}}\left[ \frac{2\pi\nu D}{\zeta}\, \min\{\gamma,\,\Delta\} \tanh \frac{\min\{\gamma,\,\Delta\}}{2T} \right]^{1/4}.
 \label{qdeveloped}
\end{equation}
As expected, $q$ grows with decreasing $\zeta$ although relatively slow.

\subsection{Helical state at the resonance}

Another interesting point on the phase diagram Fig.\ \ref{fig:helical state} that requires a special treatment is the resonance $h = \gamma$. Parameters of the Landau functional Eq.\ (\ref{Landaupsiq}) formally diverge at this point hence a simple expansion in small $q$ is invalid. We will use a more general identity (\ref{zetaq}) instead.

At zero temperature, the sum over Matsubara energies in Eq.\ (\ref{zetaq}) is replaced by an integral. At the point of resonance $h = \gamma$ and at $q = 0$ this integral is the same as in the definition of the parameter $\alpha$, Eq.\ (\ref{alpha}), and features the same logarithmic divergence, Eq.\ (\ref{alphalog}). It can be traced back to the dependence of $\sinh m$ on $\epsilon$. In the uniform magnetic state at low energies, we can use Eq.\ (\ref{m}) and approximate
\begin{equation}
    \sinh^2 m \approx \frac{1}{2} \frac{\gamma \Delta}{\gamma + \Delta} \frac{1}{\epsilon}.
    \label{mInResonanceInArbEnergies}
\end{equation}
This identity is valid provided $m \gg 1$.

The same estimate for $\sinh m$ can be derived directly from the action (\ref{Sdeveloped}). This is achieved by using the uniform solution for $\theta$ from Eq.\ (\ref{theta}) and taking the limit of large $m$. Then the action density takes the following form:
\begin{multline}
 S
  = -\pi\nu \Biggl[
      \frac{D q^2}{4} e^{2 m} \\ 
      +  4 \gamma e^{-m}\sqrt{1  + \left(\frac{(\gamma + \Delta) \epsilon}{2 \gamma \Delta} e^{2m}\right)^2} \Biggr].
 \label{Sresonance}
\end{multline}
We also retain here the $q^2$ term that is needed to limit the logarithmic divergence at low energies. To minimize this action with respect to $m$ requires solving a cubic equation. We however do not need its exact solution. It is sufficient to note that at not too small energies the solution is independent of $q$ and reproduces Eq.\ (\ref{mInResonanceInArbEnergies}). The parameter $m$ grows with decreasing $\epsilon$ until the effect of $q^2$ term kicks in. Setting $\epsilon = 0$ in the action above, we find the maximal value achieved by $m$:
\begin{equation}
    e^{3m_\text{max}} = \frac{16 \gamma}{Dq^2}.
    \label{mmax}
\end{equation}
Comparing this value with Eq.\ (\ref{mInResonanceInArbEnergies}) we establish the lower bound on the logarithmic integral in $\epsilon$. The upper bound is set at the scale where $m$ ceases to be large and the asymptotics (\ref{mInResonanceInArbEnergies}) fails. This yields the following energy window for the logarithmic integral in Eq.\ (\ref{zetaq}):
\begin{equation}
 \frac{\gamma \Delta}{\gamma + \Delta} \left( \frac{\gamma}{Dq^2} \right)^{2/3}  \lesssim \epsilon \lesssim \frac{\gamma \Delta}{\gamma + \Delta}.
 \label{Energy scales}
\end{equation}
Within this approximation, the identity (\ref{zetaq}) becomes
\begin{equation}
 \zeta
  \approx \frac{\nu D}{3} \frac{\gamma \Delta}{\gamma + \Delta} \ln \frac{\gamma}{D q^2}.
\end{equation}
The optimal value of $q$ at the point of resonance is then exponentially small:
\begin{equation}
 q
  \propto \sqrt{\frac{\gamma}{D}}  \exp\left[ -\frac{3 \zeta}{2 \nu D} \frac{\gamma + \Delta}{\gamma \Delta} \right].
 \label{qresonance}
\end{equation}
Numerical prefactor in this expression is of the order $1$. It can be also found analytically in the two limits $\gamma \gg \Delta$ and $\gamma \ll \Delta$, see Appendix \ref{Appendix: Resonanse}. 

On one hand, at the resonance, a nonuniform magnetic state emerges even in stiff ferromagnets with relatively large $\zeta$. This is the most favorable parameter range in the phase diagram to observe appearance of the nonuniform magnetization. On the other hand, the value of $q$ becomes exponentially small as $\zeta$ is increased. This is the reason why the Landau functional expansion (\ref{Landaupsiq}) fails to describe this limit.

\section{Summary and discussion}
\label{Sec:summary}

To summarize, we have investigated the interplay between ferromagnetism and superconducting proximity effect in a thin disordered ferromagnetic layer coupled to a bulk superconductor, Fig.\ \ref{fig:SF}. Our main finding is the possibility of a continuous phase transition between uniform and nonuniform magnetic state of this system. We have identified conditions under which this transition occurs depending on the parameters of the junction. In the most general case, these parameters include exchange field $h$ and stiffness $\zeta$ of the ferromagnet, order parameter $\Delta$ of the superconductor, conductance of the SF boundary $\gamma$ [measured in energy units, Eq.\ (\ref{gamma})], and temperature $T$. Phase diagrams for this transition in the space of parameters are shown in Figs.\ \ref{fig::alpha delta} and \ref{fig::alpha temperature}. Transition into a nonuniform magnetic state occurs provided the stiffness $\zeta$ of the ferromagnet is lower than a certain threshold value that depends on all other parameters.

We have constructed the Landau functional Eq.\ (\ref{Landau})---an expansion of the free energy in the vicinity of the phase transition. This functional allows for any type of magnetic texture and can be used to directly compare different types of nonuniform magnetic orders on the energy scale. An interesting feature of the Landau functional is the divergence of its parameters at the point $h = \gamma$ at zero temperature. This divergence implies that superconducting proximity effect is strongest when the coupling between normal and superconducting parts of the junction is in resonance with the ferromagnet's exchange field. Formally, a nonuniform magnetic state appears at this resonance point even for relatively high values of the magnetic stiffness.

We have further studied the simplest possible magnetic texture in the SF junction---the helical state. Near the phase transition, the most energetically favorable configuration corresponds to the magnetization rotating with a constant rate in real space (quantified by the wave vector $q$) along a great circle in the spin space. This helical state was also studied far from the transition, where $q$ is no longer a small parameter. In particular, we have found a power-law increase of $q \propto \zeta^{-1/4}$ with lowering magnetic stiffness $\zeta$, see Eq.\ (\ref{qdeveloped}). We have also studied the point of resonance $h = \gamma$, where the Landau functional expansion breaks down, and found a finite albeit exponentially small value of $q$ there, see Eq.\ (\ref{qresonance}). 

In this work, we have focused on the simplest model of the ferromagnet disregarding possible magnetic anisotropy or spin-orbit coupling. Such an idealized model has no coupling between directions in real and spin spaces hence any magnetic texture can be freely rotated as a whole in the spin space without additional energy cost. It is straightforward to include magnetic anisotropy in our model by just adding the corresponding term to the magnetic energy (\ref{Fmag}). For ferromagnets with an easy plane, such an extra term will not change our findings, it will just fix the preferred plane of rotation for the helical state. Taking into account the energy of magnetic field outside of the ferromagnetic layer (demagnetizing factor) produces a similar effect: rotation of magnetization occurs preferably in the plane of the sample. In the sample with the easy axis anisotropy, a helical state with a nontrivial aperture $\psi \neq \pi/2$ may arise. Effects of spin-orbit coupling, e.g., Dzyaloshinskii-Moriya interaction \cite{Dzyaloshinsky58, Moriya60}, can be also included directly in the magnetic free energy.

Throughout the paper, we have assumed a tunnel contact between superconductor and ferromagnetic layer. In the opposite case of a transparent boundary, the inverse proximity effect on the superconducting side of the junction should be also taken into account. However, qualitative picture of the phase transition between uniform and nonuniform magnetic states in such a junction should be similar to the one found here.

Similar effects based on the interplay between superconducting proximity and ferromagnetism will also arise in junctions of other geometries. The simplest generalization of the junction studied in this paper is an SFS heterostructure with a thin ferromagnetic layer sandwiched between two bulk superconductors. In such a junction, the strength of proximity effect can be tuned by the Josephson current applied between superconducting leads. This provides an additional control parameter to study the phase transition into a nonuniform magnetic state. Another related geometry involves a long ferromagnetic wire attached to a bulk superconducting lead only by its end. In such a system, proximity effect may induce a nonuniform magnetic order where magnetization changes its direction as a function of the distance from the superconductor. This problem will be studied in a separate publication.

Let us estimate the parameters for a feasible experimental setup to observe formation of the nonuniform magnetic state predicted in this paper. Although most ferromagnets have a relatively large exchange field $h$ much exceeding typical superconducting gap $\Delta$, it is advantageous to reduce their ratio. For instance, a dilute ferromagnetic alloy Cu$_{0.48}$Ni$_{0.52}$ has the Curie temperature of the order $20$--$30$K that is comparable to the superconducting gap of niobium $\Delta = 17$K. This combination of materials was used in Ref.\ \cite{Ryazanov01} to fabricate SFS Josephson junctions with phase shift $\pi$. These junctions also had a very short electron mean free path $l \sim 1$nm.

For the problem studied in this paper, a ferromagnetic layer of the same material with thickness $d \lesssim 10$nm would behave as a true 2D metal at energy scales of the order of $h$. Superconducting part of the junction should be relatively clean with the diffusion constant $D \gtrsim h^2 d^2/\Delta \sim 2 \cdot 10^{-3}$m$^2$/s in order to neglect possible inverse proximity effect. Magnetic stiffness of the ferromagnetic layer can be estimated from the same Curie temperature and turns out to be $\zeta \sim \nu D \Delta$. Most challenging parameter is the boundary conductance $G_T$. To make the proximity effect stronger, we would like to get as close as possible to the point of resonance $h = \gamma$. Tunneling conductance per unit area is proportional to the density of states and to the transparency $\mathcal{T}$ of the boundary. Using Eq.\ (\ref{gamma}), we can estimate the optimal transparency as
\begin{equation}
 \mathcal{T}
  \sim \frac{h d}{v_F}
  \sim 0.1.
\end{equation}
This value corresponds to a relatively good electric contact between superconductor and ferromagnetic layer. We conclude that observation of the transition into a nonuniform magnetic state is quite feasible in a properly constructed SF junction.

\acknowledgments{We are grateful to M.\ Feigel'man, Ya.\ Fominov, M.\ Ismagambetov and A.\ Lunkin for fruitful discussions. We also thank A.\ Levina for helping with the design of Figs.\ \ref{fig:SF} and \ref{fig:Helical state demonstration}}.

\appendix
\section{Wave vector at resonance}
\label{Appendix: Resonanse}

As was discussed earlier, in the resonant regime $h = \gamma$ at zero temperature the energy sum in Eq.\ (\ref{zetaq}) is logarithmically large and accumulates over a wide range of energies, see Eq.\ (\ref{Energy scales}). The parameter $m$ also varies from $0$ up to a large value $m_\text{max}$ at $\epsilon = 0$, see Eq.\ (\ref{mmax}). Our goal in this Appendix is to find the wave vector $q$ at the resonance up to the numerical prefactor. In order to do this, we introduce an intermediate energy scale $\Lambda$ and split the energy integral in Eq.\ (\ref{zetaq}) into upper and lower parts. The corresponding threshold value of $m$ we will denote $\mu = m(\Lambda, q)$.
\begin{subequations}
    \begin{gather}
        \frac{\gamma \Delta}{\gamma + \Delta} \left( \frac{\gamma}{Dq^2} \right)^{2/3}  \ll \Lambda \ll \frac{\gamma \Delta}{\gamma + \Delta} \label{lambda} \\
        1 \ll \mu \ll m_\text{max}.
    \end{gather}
    \label{Lambda and mu}
\end{subequations}

In the upper region $\epsilon \geq \Lambda$, we can neglect the inhomogeneity term $Dq^2$ in the action density (\ref{Sdeveloped}) and use the values of $\theta$ and $m$ from Eqs.\ (\ref{mtheta}). In the lower region $\epsilon \leq \Lambda$, the value of $m \geq \mu$ is large and the action can be approximated by Eq.\ (\ref{Sresonance}). In the intermediate region near the energy $\epsilon \sim \Lambda$, both approximations are valid and provide the same estimate (\ref{mInResonanceInArbEnergies}) for the value of $m$. 
This establishes a relation between $\Lambda$ and $\mu$:
\begin{equation}
 \Lambda = \frac{2\gamma \Delta}{\gamma + \Delta}\, e^{-2\mu}.
\end{equation}

We are now ready to estimate the integral in Eq.\ (\ref{zetaq}) in the two asymptotic limits. In the region $\epsilon \leq \Lambda$, the value of $m(\epsilon,q)$ is determined by minimizing the action Eq.\ (\ref{Sdeveloped}), which is equivalent to solving a cubic equation. We can overcome this obstacle by changing the integration variable from $\epsilon$ to $m$:
\begin{multline}
 \int_0^\Lambda d\epsilon \: \sinh^2 m(\epsilon, q)
  = \epsilon \sinh^2 m(\epsilon, q) \bigg|_{\epsilon = 0}^{\epsilon = \Lambda} \\
    + \int_{\mu}^{m_\text{max}} dm\, \epsilon(m, q) \sinh 2m.
    \label{LowEnergyAlphaIntegral}
\end{multline}
Here the function $\epsilon(m,q)$ is the result of minimization of Eq.\ (\ref{Sdeveloped}) with respect to $m$ and then resolving the corresponding equation in $\epsilon$. Using the relation (\ref{mmax}), we can express $q$ in terms of $m_\text{max}$. Then the integrand in Eq.\ (\ref{LowEnergyAlphaIntegral}) takes a simple form in terms of the new variable $\xi = e^{3(m - m_\text{max})}$
\begin{equation}
 \epsilon \sinh(2m)
  = \frac{\gamma \Delta}{\gamma + \Delta} \sqrt{1 + 2 \xi^2 - 2 \xi \sqrt{2 + \xi^2}}.
\end{equation}
Integration in the lower energy region then yields
\begin{multline}
 \int_0^\Lambda d\epsilon \: \sinh^2 m(\epsilon, q) \\
   = \frac{\gamma \Delta}{\gamma + \Delta} \left[ -\mu - \frac{1}{3} \ln \frac{Dq^2}{\gamma} + \frac{3}{2} \ln 2 + \frac{1}{6} - \frac{\pi}{12} \right].
   \label{ResAlphaLower}
\end{multline}

Now let us estimate the integral (\ref{zetaq}) in the upper energy region. In this regime we can neglect the wave vector $q$ altogether, and take the value of $m$ directly from Eq.\ (\ref{m}). The resulting integral is very complicated but greatly simplifies in the two limits of large and small $\gamma$:
\begin{equation}
    \int_\Lambda^\infty d\epsilon\, \sinh^2 m(\epsilon)
    = \frac{\gamma \Delta}{\gamma + \Delta} \begin{dcases}
        \mu + \frac{1}{2}\ln 2 -1, &\gamma \ll \Delta, \\
        \mu - \frac{1}{2}, &\Delta \ll \gamma.
    \end{dcases}
    \label{ResAlphaUpper}
\end{equation}

Complete integral over all energies in Eq.\ (\ref{zetaq}) is the sum of the two contributions Eqs.\ (\ref{ResAlphaLower}) and (\ref{ResAlphaUpper}). We observe that in this sum, the parameter $\mu$ cancels out. This happens for any ratio $\gamma/\Delta$ since the whole integral does not depend on the artificially introduced splitting parameter.

Using Eq.\ (\ref{zetaq}), we obtain a more accurate version of the result (\ref{qresonance}) for the wave vector $q$ at the resonance:
\begin{equation}
    q = C \sqrt{\frac{\gamma}{D}} \exp{\left(-\frac{3}{2}\frac{\zeta}{\nu D} \frac{\Delta + \gamma}{\Delta \gamma} \right)}.
\end{equation}
Here the prefactor $C$ takes the following values in the limiting cases:
\begin{equation}
     C = \begin{dcases}
         8 \exp \left( - \frac{10 + \pi}{8} \right) \approx 1.55, &\gamma \ll \Delta, \\
         2^{9/4} \exp \left( -\frac{4 + \pi}{8} \right) \approx 1.95, &\gamma \gg \Delta, \\
         \approx 1.52\: \text{(minimum)}, &\gamma \approx 1.67 \Delta .
     \end{dcases}
     \label{C}
\end{equation}
Numerical evaluation of the integral (\ref{ResAlphaUpper}) provides the value of $C$ for an arbitrary ratio $\gamma/\Delta$. It yields the minimal value cited in Eq.\ (\ref{C}) and also shows that $C$ always stays within the interval $[1.52,\, 1.95]$.

\bibliography{bibliography.bib}

\end{document}